\documentclass[11pt]{article}
\usepackage{amsmath}
\usepackage{amssymb}
\usepackage{epsfig}
\newlength{\bredde}
\def\slash#1{\settowidth{\bredde}{$#1$}\ifmmode\,\raisebox{.15ex}{/}
\hspace*{-\bredde} #1\else$\,\raisebox{.15ex}{/}\hspace*{-\bredde} #1$\fi}
\newcommand{\be}{\begin{equation}}
\newcommand{\ee}{\end{equation}}
\newcommand{\bea}{\begin{eqnarray}}
\newcommand{\eea}{\end{eqnarray}}
\newcommand{\nn}{\nonumber}

\newcommand{\la}{\lambda}
\newcommand{\La}{\Lambda}

\newcommand{\eins}{\leavevmode\hbox{\small1\kern-3.8pt\normalsize1}}
\newcommand{\one}{\mbox{\bf 1}}
\newcommand{\e}{\mbox{e}}
\newcommand{\erf}{\mbox{erf}}
\newcommand{\erfc}{\mbox{erfc}}

\newcommand{\sect}[1]{\setcounter{equation}{0}\section{#1}}

\def\Tr{{\mbox{Tr}}}

\def\Pf{\mbox{Pf}}

\def\hx{\hat{x}}
\def\hy{\hat{y}}
\def\hz{\hat{z}}
\def\hm{\hat{m}}

\def\ha{\hat{a}}

\begin{document}
\topmargin -1.4cm
\oddsidemargin -0.8cm
\evensidemargin -0.8cm
\title{\Large{{\bf
Individual Eigenvalue Distributions for the Wilson Dirac Operator
}}}

\vspace{1.5cm}
\author{~\\{\sc G.~Akemann}$^1$ and {\sc A.~C.~Ipsen}$^{2}$
\\~\\$^1$Department of Physics,
Bielefeld University,
Postfach 100131,
D-33501 Bielefeld, Germany
\\~\\
$^2$Niels Bohr International Academy and Discovery Center,
Niels Bohr Institute,\\
Blegdamsvej 17, DK-2100 Copenhagen {\O}, Denmark}

\date{}

\maketitle
\vfill
\begin{abstract}
We derive the distributions of individual eigenvalues for the Hermitian
Wilson Dirac Operator $D_5$ as well as for real eigenvalues of the
Wilson Dirac Operator $D_W$. The framework we provide is valid in the
epsilon regime of chiral perturbation theory for any number of
flavours $N_f$ and for non-zero low energy constants $W_{6,7,8}$. It
is given as a perturbative expansion in terms of the $k$-point
spectral density correlation functions and integrals thereof, which in
some cases reduces to a Fredholm Pfaffian. For the real eigenvalues of
$D_W$ at fixed chirality $\nu$ this expansion truncates after at most
$\nu$ terms for small lattice spacing $a$. Explicit examples for the
distribution of the first and second eigenvalue are given in the
microscopic domain as a truncated expansion of the Fredholm Pfaffian
for quenched  $D_5$, where all $k$-point densities are explicitly
known from random matrix theory. For the real eigenvalues of quenched
$D_W$ at small $a$ we illustrate our method by the finite expansion of
the corresponding Fredholm determinant of size $\nu$.

\end{abstract}


\vfill

\thispagestyle{empty}
\newpage

\renewcommand{\thefootnote}{\arabic{footnote}}
\setcounter{footnote}{0}

\sect{Introduction}\label{intro}

At low energies Quantum Chromodynamics (QCD) 
has a very convenient analytical testing ground,
the so-called epsilon regime of chiral perturbation theory (XPT) \cite{GL}.
In this limit standard XPT is reduced to a group integral
over the zero-momentum modes that are then non-perturbative and dominate
the theory, times a Gaussian free field theory of propagating Goldstone
bosons. Although being an unphysical limit the analytic predictions in this
regime are useful and can be tested against numerical results from lattice gauge
theory.
Of particular interest for the establishment of chiral symmetry
breaking and the effect of gauge field topology is the Dirac operator
spectrum. There are several equivalent methods to compute correlation
functions of Dirac operator eigenvalues in this regime.  Either one
starts from XPT in a 
partially quenched also called graded setting which is based on supergroup
integrals, see e.g. \cite{DOTV}, or equivalently one can 
use replicated partition
functions \cite{DS00}. Yet another method that was
actually first proposed in \cite{SV} uses chiral Random Matrix Theory
(RMT), being equivalent to the epsilon regime of XPT at leading order (LO)
\cite{DOTV,TV}. This latter approach is
best suited to compute the distribution of individual
Dirac operator eigenvalues \cite{DNW,DN} - which in turn are particularly clean
objects for a comparison with lattice data.

In the continuum limit this setup has already been very useful,
in order to determine the low energy constants (LEC) in the chiral
Lagrangian or to
test lattice algorithms for their topology dependence, and we refer to
\cite{Poulrev} for a recent review including references. 
In XPT there are only two LEC to LO, the chiral condensate $\Sigma$
which is the order parameter of chiral symmetry breaking, and the
Pion decay constant $F$. Individual Dirac operator eigenvalues allow
for a very detailed error analysis in the determination of $\Sigma$
compared to 
the average density through the Banks-Casher relation, and we refer 
e.g. to \cite{Bernardoni} for recent work. In order to
make the epsilon regime also sensitive to $F$ a coupling to imaginary
iso-spin chemical potential $\mu$ has been proposed and realised in
\cite{Imiso}. This leads to realistic values of $F$ when computing and
optimising higher order corrections \cite{CT}. 
The computations were done in a partially quenched setting, where a
$\mu$-dependent Dirac 
operator is probed with $\mu$-independent configurations. A wealth of
results exists for the
density correlation functions with \cite{Imiso,ADOS,CT} and without
$\mu$ \cite{SV,RMT2},
as well as for individual eigenvalue distributions with \cite{AD08,AI}
and without
$\mu$ \cite{DNW,DN}.

What happens if one wants to include the effect of finite lattice
spacing $a$, which is unavoidably present in all simulations? Perhaps
surprisingly the effect is much simpler in Wilson XPT developed in
\cite{SharpeSingleton,Sharpe},
than in its staggered version \cite{James}.
This is despite the fact that
the Wilson term completely breaks chiral symmetry whereas the
staggered Dirac operator partly preserves it.
In a Symanzik type expansion Wilson XPT requires three extra LEC
$W_{6,7,8}$ to LO, that have to be determined in order to quantify the effect of
a finite lattice spacing\footnote{This is so even though they
  represent unphysical
lattice artefacts, with their values depending on the lattice action
considered.}, and we refer to \cite{GS} for review articles.

The epsilon regime with Wilson fermions was put into focus in
\cite{BNS}, addressing mesonic correlation functions.
Then two years ago the microscopic spectral density of the Wilson
Dirac operator in the epsilon regime was computed \cite{DSV}.
The study of this operator
has since seen a remarkably fast progress. First of all one has to
distinguish between the non-Hermitian Wilson Dirac operator $D_W$ and
its Hermitian version $D_5$, and their corresponding densities.
While we refer to ref. \cite{ADSV} for a discussion of the relation
between topology and chirality 
$\nu$ at finite-$a$, and to ref. \cite{KVZ} for the
distinction of left and right eigenvalues for $D_W$, let us give a
short list of what is known to date about the eigenvalue correlation
functions.

The quenched microscopic density for $D_5$ and for the real
eigenvalues of $D_W$ 
were computed in \cite{DSV,ADSV}, respectively. These
results were extended to $N_f=1$ flavours in \cite{ADSVLatNf,RNL}
and to $N_f=2$ in \cite{SV2011}. There, also the
quenched two-point eigenvalue density was given explicitly, and a computational
scheme for higher densities and $N_f>2$ was set up although it rapidly
becomes cumbersome. In \cite{AN} all quenched higher density
correlations were computed for $D_5$ at $\nu=0,1$. The corresponding
results for the complex density of $D_W$ eigenvalues including all
$\nu$ were given 
in \cite{KVZ}, where further results to all orders were announced
which have just appeared \cite{M11}.
Most of the previous results for the density (except \cite{RNL}) were given for
$W_8\neq0$ and $W_{6,7}=0$, which is expected to be a good
approximation for a large number of colours \cite{KL}. However, it was
also understood in 
\cite{ADSV} how to derive $W_{6,7}\neq0$ results from $W_{6,7}=0$
results 
on the level of XPT 
by one extra Gaussian integration per LEC. 
This was now extended to an RMT setting in \cite{M11}.
Furthermore, in
\cite{ADSV} it was shown that in the epsilon regime QCD inequalities
lead to a restriction on the sign of $W_8$ to be positive for
$W_{6,7}=0$. 
This constraint was very recently extended to different combinations of LEC
\cite{HS1,HS,KSV}.
We note that the signs (of combinations) of
these LEC are not merely of academic interest, but decide crucially on
the scenarios of possible phase transitions, see e.g. the discussion
in \cite{Sharpe,HS1,KSV}. 
Finally we would also like to mention the computation of the mean
spectral density of $D_5$ in \cite{NS}.

Based on previous experience at $a=0$ how useful
individual eigenvalue distributions are, in \cite{DHS,DWW} first
lattice simulations to compare to the above epsilon regime predictions at
$a\neq0$ were made. There the individual eigenvalue distributions were
generated {numerically} from RMT. We fill this gap here by
providing the framework of how to analytically
compute these distributions. This setup holds
for general $N_f$ and non vanishing LEC $W_{6,7,8}$.  
Our
explicit examples  are all quenched and at $W_{6,7}=0$, but 
our framework
can be used for all higher density correlations for $D_5$ and $D_W$ at
$N_f>0$ that are available.

The content of this paper is organised as follows. In section
\ref{general} we set up the general framework of how to compute
individual eigenvalue distributions or their cumulative distributions
in terms of density correlation functions which we assume to be
given. In section \ref{DW5} we 
then specify the joint probability distributions for $D_5$-
(subsect. \ref{jpdfD5}) and real $D_W$-eigenvalues
(subsect. \ref{jpdfDW})
based on RMT. At $W_{6,7}=0$ this leads to a closed Fredholm Pfaffian
expression for 
the former, and a distribution of finitely many $\nu$ real eigenvalues
at small spacing $a$ for the latter.
Section \ref{examples} is devoted to explicit examples in the
microscopic or epsilon regime. For quenched $D_5$
we give the first and second eigenvalue in a truncated
expansion of the Fredholm Pfaffian  (subsect. \ref{ex5}).
In the case of $D_W$ (subsect. \ref{exW}) the approximate 
distributions are given in a finite series of expansion terms
for $\nu=2,3$.
Our conclusions are summarised in section \ref{conc}.

\sect{Individual eigenvalue distributions: general theory}\label{general}

In this section we describe the general formalism how to compute the
distribution of individual eigenvalues and their cumulative
distribution, having the Wilson Dirac operator and its Hermitian
counterpart in mind. 
We will focus on real eigenvalues here, although this setup can also
be generalised to complex eigenvalues after choosing a family of
contours to order them, and we refer to \cite{ABPS} for details in the
case of a real chemical potential.

Our starting point is chosen to be quite broad, given in terms of a
partition function and its joint probability distribution (jpdf) of all
eigenvalues that is invariant under permutations of eigenvalues,
without further symmetry requirements. In particular we do not require
the jpdf to be a determinant or Pfaffian in this section. 
As an example the jpdf can be obtained from the corresponding random
two-matrix models 
and their representation in terms of $D_5$- or real $D_W$-eigenvalues,
which we will specify in section \ref{DW5}. 
Note that when introducing $W_{6,7}\neq0$ by extra Gaussian integrations
\cite{ADSV,KSV,M11} the resulting jpdf is no longer a determinant or
Pfaffian. Therefore it is important to be more general here. 
Our final expressions for individual eigenvalues will only contain
$k$-point spectral densities and their integrals.
For that reason this setup also applies to partially quenched XPT
where these densities can be generated by
introducing pairs of fermionic and bosonic source terms. The analogous
statement for spacing $a=0$ was reported in \cite{AD03}.

We begin by defining the partition function of the theory
under consideration,
\be
{\cal Z}\ \equiv\ \left(\int_{-\infty}^{\infty}\right)^n
\prod_{i=1}^nd\la_i\ {\cal P}_{jpdf}(\la_1,\ldots,\la_n) \ .
\label{Zjpdf}
\ee
It is given by $n$ real integrals over the so-called joint
probability distribution function ${\cal P}_{jpdf}$, which
is taken to be invariant under permutations of all eigenvalues,
without further symmetry requirement. 
It may depend on external parameters such as quark masses, the number of
flavours or the chirality.
We also define the $k$-point density correlation function,
\be
R_k(\la_1, \ldots, \la_k) \ \equiv\ \frac{n!}{(n-k)!} \frac{1}{{\cal Z}}
\left(\int_{-\infty}^\infty\right)^{n-k} d\la_{k+1} \ldots d\la_n
\,{\cal P}_{jpdf}(\la_1, \ldots, \la_n)\ ,
\label{Rk}
\ee
normalised by the partition function. The simplest example is just the
spectral density for $k=1$. For $k=n$ the density is simply the
normalised jpdf itself, 
$R_n={\cal P}_{jpdf}/{\cal Z}$.
The $R_k$ are supposed to be known in the following as all
other quantities will be expressed in terms of them.

The $k$-th gap probability or cumulative distribution is defined as
follows:
\be
E_k(b,c) \ \equiv\ \frac{n!}{(n-k)!}\frac{1}{{\cal Z}}
\left(\int_b^c\right)^k d\la_1\ldots d\la_{k}
\left(\int_{-\infty}^b
+\int_c^\infty\right)^{n-k} d\la_{k+1}\ldots d\la_n
\,{\cal P}_{jpdf}(\la_1, \ldots, \la_n) \ ,
\label{Ek}
\ee
for $k=0,1,\ldots,n$.
It is proportional to the probability of having exactly $k$ out of $n$
eigenvalues inside the interval $[b,c]$, and $n-k$ eigenvalues outside
this interval. Note that we do not specify any ordering among the $k$
or $n-k$ eigenvalues respectively, nor do we specify how many of the
$n-k$ eigenvalues are less than $b$ or larger than $c$ (we assume
$b<c$ without loss of generality). The simplest example with $k=0$
gives the probability of having $[b,c]$ empty of eigenvalues.

A remark is in order here. 
In the simplest jpdf we have in mind
here that describes the transition from the Gaussian Unitary Ensemble
(GUE) to the chiral GUE in RMT, the eigenvalues live on the 
full real line $\mathbb R$ and cannot be restricted to the positive half line
${\mathbb R}_+$. In particular there is no chiral symmetry as for the
chiral GUE, that for
every eigenvalue $\la_j>0$ there is also an eigenvalue at $-\la_j<0$.
Only on  average the distributions of individual eigenvalues may be
symmetric with respect to the origin in some special 
case ($\nu=0$). Apart from that the derivation here is only a mild
generalisation of \cite{AD03} which we follow closely.
Note also that we will still count the eigenvalues from the
origin onwards, e.g. by choosing $[b,c]=[0,c]$ for the positive
eigenvalues, 
as this is where the
microscopic large-$n$ limit will be taken eventually.

Rewriting $\int_{-\infty}^b+\int_c^\infty= \int_{-\infty}^\infty-\int_b^c$,
applying the binomial formula and using the permutation invariance of
the jpdf we can rewrite the $k$-th gap probability as
\bea
E_k(b,c) &=& \frac{n!}{(n-k)!}\frac{1}{{\cal Z}}
\left(\int_b^c\right)^kd\la_1\ldots d\la_{k} \nn\\
&&\times\sum_{l=0}^{n-k} (-1)^l {n-k \choose l}
\left(\int_{-\infty}^\infty\right)^{n-k-l}
\left(\int_b^c\right)^l d\la_{k+1}\ldots d\la_n \,{\cal P}_{jpdf}(\la_1,
\ldots, \la_n)\nn\\
&=& \sum_{l=0}^{n-k} (-1)^l \frac{1}{l!}
\left(\int_b^c\right)^{k+l}  d\la_1\ldots d\la_{k+l}\ R_{k+l}( \la_1,\ldots,
\la_{k+l})\ ,
\label{EkRk}
\eea
where for $k=l=0$ the first term in the last sum is unity. For the simplest
example $k=0$ we thus have
\bea
E_{k=0}(b,c)&=& 1-\int_b^c d\la_1\, R_1(\la_1)+\frac12
\int_b^c \int_b^cd\la_1d\la_2\,
R_2(\la_1,\la_2)- \frac{1}{6}\left(\int_b^c \right)^3d\la_1d\la_2d\la_3
R_3(\la_1,\la_2,\la_3) \nn\\
&&+\
\ldots ~,
\label{E0example}
\eea
where we only show the first few terms in the sum containing $n+1$
terms. Note however 
that when considering the number of real eigenvalues of $D_W$, we can
set to a good approximation 
$n=\nu$ at small $a$, and then we will only need a small and fixed number of
terms. In particular for 
$n=3$ eq. (\ref{E0example}) would be exact, and for $n=2$ we only have
the first 3 terms\footnote{For $n=1$ with a single eigenvalue
no new information is gained from $E_0$ compared to the density itself, see also
eq. (\ref{p1example}) for the first eigenvalue.}.
For the Hermitian Wilson Dirac operator $D_5$, $n$
will become proportional to the number of eigenvalues and thus will be
taken to infinity. Even in that case taking only the first few terms in
the now infinite series eq. (\ref{EkRk}) will yield an excellent
approximation, as we 
will see in section \ref{examples} and as was known already for $a=0$
\cite{AD03}. 

The $k$-th gap probability can be most conveniently written in terms
of a generating function,
\be
E(b,c;\xi)\ \equiv\  1 + \sum_{l=1}^{n} (-\xi)^l \frac{1}{l!}
\left(\int_b^c\right)^l  d\la_1\ldots d\la_{l}\ R_{l}(\la_1,\ldots, \la_{l}) \ .
\label{Egen}
\ee
We then have
\be
E_k(b,c)\ =\ (-1)^{k} \left.\frac{\partial^{k}}{\partial\xi^{k}} \
E(b,c;\xi) \right|_{\xi=1} \ , \ \ \mbox{for}\ k=0,1,\ldots,n\ \ .
\label{Ekpartial}
\ee
The probability $p_k(b,c)$ of having one eigenvalue at the upper
boundary $\la=c$, with
$k-1\geq0$ eigenvalues in $[b,c]$ and all other eigenvalues outside this
interval is defined as
\bea
p_k(b,c) &\equiv& k {n \choose k}\frac{1}{{\cal Z}}
\left(\int_b^c\right)^{k-1}d\la_1\ldots d\la_{k-1}
\left(\int_{-\infty}^b+\int_c^\infty\right)^{n-k} d\la_{k+1}\ldots d\la_n \nn\\
&&\times
{\cal P}_{jpdf}(\la_1, \ldots,\la_{k-1},\la_k=c,\la_{k+1},\ldots, \la_n)\ ,
\label{pk}
\eea
for $k=1,2,\ldots,n$.
For example if we choose $b=0$ and $k=1$ this gives the distribution
of the first positive eigenvalue on ${\mathbb R}_+$. The $p_k(b,c)$
can be obtained by differentiation 
from the corresponding gap probability
as we will show below. Analogously
we can  define the probability $q_k(b,c)$ of having one eigenvalue at
the lower boundary $\la=b$, with
$k-1\geq0$ eigenvalues in $[b,c]$ and all other eigenvalues outside this
interval:
\bea
q_k(b,c) &\equiv& k {n \choose k}\frac{1}{{\cal Z}}
\left(\int_b^c\right)^{k-1}d\la_1\ldots d\la_{k-1}
\left(\int_{-\infty}^b+\int_c^\infty\right)^{n-k} d\la_{k+1}\ldots d\la_n \nn\\
&&\times
{\cal P}_{jpdf}(\la_1, \ldots,\la_{k-1},\la_k=b,\la_{k+1},\ldots, \la_n)\ ,
\label{qk}
\eea
for $k=1,2,\ldots,n$. This definition is convenient when counting
negative eigenvalues from the origin onwards. For example 
the first negative eigenvalue on ${\mathbb R}_-$ follows by differentiating
$E_0(b,c)$ with respect to $b$ and by choosing $b<0=c$.

The explicit relation between individual eigenvalue distributions 
and gap probabilities are as follows, where we restrict ourselves to
the $p_k(b,c)$. The relations for the $q_k(b,c)$ are similar. We obtain
from $E_k(b,c)$ by differentiation
\be
\partial_c E_k(b,c) = k!\, \left( p_{k}(b,c) -
p_{k+1}(b,c)\right)\ ,
\label{Ekpk}
\ee
(with $p_0(b,c)\equiv0$) or, after taking the sum on both sides,
\be
p_{l}(b,c)\ =\ - \sum_{k=0}^{l-1}\frac{1}{k!}\partial_c E_k(b,c)\ .
\label{pkEk}
\ee
In the case of the simplest example $k=1$ and $b=0$ 
the expansion eq. (\ref{E0example}) maps to
\be
p_1(0,c)\ = \ -\frac{\partial}{\partial c} E_0(0,c)
\ =\ R_1(c) -\int_0^cd\la\, R_2(\la,c) \ + \ldots \ ,
\label{p1example}
\ee
for the distribution of the first positive eigenvalue on ${\mathbb R}_+$
after choosing $b=0$,
and
\be
p_2(0,c)\ = \ \int_0^cd\la\, R_2(\la,c) \ +\ \ldots \ \ .
\label{p2example}
\ee
for the second positive eigenvalue. Note again that if we had $n=2$
here, the expressions in eqs. (\ref{p1example}) and
(\ref{p2example}) would be exact, without further corrections.

\sect{Application to the Wilson Dirac operators $D_5$ and $D_W$
}\label{DW5}

In this section we specify to the two RMT for the Wilson Dirac
operator $D_W$ and its Hermitian counterpart $D_5$ for an arbitrary
number of 
flavours $N_f$, and give their respective jpdf. 
In the limit of large matrices specified in the next section
\ref{examples} they describe the epsilon regime of Wilson XPT.

\subsection{The jpdf and quenched $k$-point densities of $D_5$}\label{jpdfD5}
We begin with the RMT for the Hermitian Wilson Dirac operator with
$W_{6,7}=0$ in the 
form as it was introduced in \cite{AN}, apart from putting an explicit
$n$-dependence into the weight function here. The partition function
is 
defined as
\bea
{\cal Z}_5(m;z)&\equiv&
\int dH \int dW \prod_{f=1}^{N_f}
\det[D_5 +z_f\one_N]
\exp\left[-\ \frac{n}{(1-a^2)}\Tr WW^\dag-\ \frac{n}{2a^2}\Tr H^2\right],
\label{Z5}\\
D_5 &\equiv&
\left(\begin{array}{cc}
m\one_n& W\\
W^\dag&-m\one_{n+\nu}\\
\end{array}\right)
\ +  \ H\ ,
\label{D5}
\eea
where we have introduced the Hermitian random matrix\footnote{Note that
  in the original proposal \cite{ADSV} 
  instead of a full matrix $H$ only two Hermitian matrices were added
  on the diagonal blocks, similar to eq. (\ref{DW}).} $H=H^\dag$ of
size $N\times N$ with $N=2n+\nu$ and the complex random matrix
$W\neq W^\dag$ of size
$n\times(n+\nu)$. Here $\nu\in\mathbb N$ will be kept fixed and of
order one, it plays the r\^ole of chirality \cite{ADSV}.
The parameter $a^2\in[0,1]$ incorporates the influence of the lattice
spacing times the LEC, $a_{lat}^2W_8$,
see eq. (\ref{micro2}) for the precise mapping between RMT and XPT
quantities. It also allows to interpolate 
between the chiral GUE ($a=0$) and the GUE ($a=1$) in RMT when setting
$N_f=m=0$. The real parameters $m$ and $z_f$ denote the standard quark
mass and sources for $\bar{\psi}\gamma_5\psi$, respectively.
The jpdf for the eigenvalues $x_j$ of $D_5$
was derived in \cite{AN} to where we refer for details,
and we only give the result
valid up to an overall ($a$-dependent) constant:
\bea
{\cal P}_5(\{x\};m;z)&\equiv&
\left(\prod_{j=1}^N
\exp\left[{-\frac{nx_j^2}{2a^2}}\right]\prod_{f=1}^{N_f}(x_j+z_f)\right)
\exp\left[{\frac{-nm^2(-2n+N(1-a^2))}{2a^2(1-a^2)}}\right]
\nn\\
&&\times\Delta_{2n+\nu}(\{x\})\ 
\Pf_{1\leq i,j\leq N;\,1\leq q\leq \nu}
\left[
\begin{array}{ll}
F(x_j-x_i;m)& \!x_i^{\, q-1}\e^{-\frac{nx_im}{a^2}}\\
 -x_j^{\, q-1}\e^{-\frac{nx_jm}{a^2}} & {\mathbf 0}_{\nu\times\nu}\\
\end{array}
\right],\nn\\
{\cal Z}_5(m;z) &=&
\left(\prod_{j=1}^{2n+\nu} \int_{-\infty}^\infty dx_j\right)
     {\cal P}_5(\{x\};m;z) \ .
\label{Z5jpdf}
\eea
Here $\Delta_N(\{x\})=\prod_{i>j}^n(x_i-x_j)$ denotes the standard
Vandermonde determinant. We have also defined the antisymmetric weight
function
\bea
F(x;m)
&\equiv&\exp\left[{\frac{nx^2(1-a^2)}{4a^2}}\right]\nn\\
&&\times\left[\erf\Big(\frac{x\sqrt{n(1-a^2)}}{2a}
+\frac{m\sqrt{n}}{\sqrt{a^2(1-a^2)}}\Big)  +
\erf\Big(\frac{x\sqrt{n(1-a^2)}}{2a}-
\frac{m\sqrt{n}}{\sqrt{a^2(1-a^2)}}\Big)  
\right],\ \ \ \ 
\label{Fdef}
\eea
appearing inside the Pfaffian. It could also be written in terms of
the generalised incomplete error function $\erf(x,y)$ (c.f. \cite{KVZ}). 
Note that the Jacobian from the diagonalisation of $D_5$ is not a
simple Vandermonde determinant to some integer power, as it is in standard
one-matrix models. The additional Pfaffian determinant containing part
of the weight function $F(x;m)$ is a feature shared by the partition function
for the non-Hermitian Wilson Dirac operator eq. (\ref{ZW}) \cite{KVZ}, as well
as by other non-Hermitian RMT, see \cite{AKP}.

The problem of computing individual eigenvalue distributions in terms
of densities is now specified. 
All $k$-point
density correlation functions have been computed explicitly in the quenched
approximation $N_f=0$ for $\nu=0,1$ \cite{AN}. For higher $\nu>1$ for
$N_f=0,1$ and $N_f=2$ 
the spectral density has been computed
explicitly
in \cite{ADSV,ADSVLatNf,RNL,SV2011}, 
respectively. There, the supersymmetric or
graded eigenvalue method was used starting from the partially quenched
chiral Lagrangian, but the results agree because of universality.
Unfortunately from the density alone
being the first term
in eq. (\ref{p1example}) we do not gain any new information regarding
the first eigenvalue.
Let us emphasise that this lack of present information is not a
restriction of our method. 
With the more general results for 
all $k$-point density correlation functions for arbitrary $N_f$ and
$\nu$ 
now at hand \cite{M11}, 
these can be simply inserted into our setup without
modifications.

We continue to recall 
some of the results of \cite{AN}. For $N_f=0$ and $\nu=0,1$ all
$k$-point density correlation functions can be written as the Pfaffian
of a $2\times2$ matrix valued kernel $K_n(x,y)$,
\be
R_k(x_1,\ldots,x_k)= \Pf_{1\leq i,j\leq k}[K_n(x_i,x_j)]\ ,\ \
K_n(x_i,x_j)=\left(
\begin{array}{cc}
I_n(x_i,x_j)& S_n(x_i,x_j)\\
-S_n(x_j,x_i)& -D_n(x_i,x_j)\\
\end{array}
\right).
\label{RkPf}
\ee
For finite-$n$ and $\nu=0$ its matrix elements read
(we refer to \cite{AN} for details, including $\nu=1$):
\bea
S^{\nu=0}_n(x,y)&=& \sum_{j=1}^n\frac{\e^{-\frac{ny^2}{2a^2}}}{r_{j-1}}
\Big(\phi_{2j-2}(x)R_{2j-1}(y)-\phi_{2j-1}(x)R_{2j-2}(y)
\Big)\ ,\nn\\
D^{\nu=0}_n(x,y)&=& \sum_{j=1}^n\frac{\e^{-n\frac{x^2+y^2}{2a^2}}}{r_{j-1}}
\Big(R_{2j-2}(x)R_{2j-1}(y)-R_{2j-1}(x)R_{2j-2}(y)
\Big)\ ,\nn\\
I^{\nu=0}_n(x,y)&=&- \sum_{j=1}^n\frac{1}{r_{j-1}}
\Big(\phi_{2j-2}(x)\phi_{2j-1}(y)-\phi_{2j-1}(x)\phi_{2j-2}(y)
\Big)\ -\ F(x-y;m)\ ,
\label{3kernels}
\eea
with the corresponding skew-orthogonal polynomials
and their integral transforms being given by
\bea
R_{2j}(x)&=&\frac{j!\ (1-a^2)^j\sqrt{2n}}{(-1)^j\sqrt{\pi}}
\int_{-\infty}^\infty ds\ \e^{-2ns^2}L_j\left(
  n\frac{(x+2ias)^2-m^2}{(1-a^2)}\right) , \label{sOPs}\\
R_{2j+1}(x)
&=&\frac{j!\ (1-a^2)^j\sqrt{2n}}{(-1)^j\sqrt{\pi}}
\int_{-\infty}^\infty ds\ \e^{-2ns^2}\sqrt{n}(x+2ias)L_j\left(
  n\frac{(x+2ias)^2-m^2}{(1-a^2)}\right),
\nn\\
\phi_j(x)&=&\int_{-\infty}^\infty dy\ \e^{-\frac{ny^2}{2a^2}}F(x-y;m)R_j(y)\ .
\label{trafo}
\eea
In particular the jpdf, eq. (\ref{RkPf}) for $k=2n+\nu$, can be written as a
single Pfaffian for both $\nu=0,1$:
\be
{\cal P}_{5}(x_1,\ldots,x_{2n+\nu};m;z=0) =
{\cal Z}_5(m;z=0)\,\Pf_{1\leq i,j\leq 2n+\nu}[K_n(x_i,x_j)]\ ,
\label{P5Pf}
\ee
which extends to $\nu>1$ \cite{M11}.
Inserting eq. (\ref{RkPf}) into the generating function for all gap
probabilities eq. (\ref{Egen}) it becomes a Fredholm Pfaffian
\cite{Rains} and can thus 
be written in a closed form
\bea
E(b,c;\xi) &=&
1 + \sum_{l=1}^{n} (-\xi)^l \frac{1}{l!}
\int_b^c  dx_1\ldots dx_{l}\ \Pf_{1\leq i,j\leq l}[K_n(x_i,x_j)] \nn\\
&=&
\prod_{j=1}^n (1-\xi\La_j) \equiv \Pf[1-\xi K]\ .
\label{EPf}
\eea
Here the $\La_j$ are the eigenvalues of the matrix integral equation
$\La f(x)=\int_b^c dy K_n(x,y)f(y)$. 
We will come back to the perturbative evaluation of this Fredholm
Pfaffian after taking the microscopic large-$n$ limit in 
section \ref{examples}. 



Finally we would like to discuss the inclusion of $a^2_{lat}W_{6,7}\sim
a^2_{6,7}\neq0$, 
following \cite{ADSV}. There, it was first
understood how to include these terms on the level of the XPT
partition function, by introducing two extra Gaussian integrations and shifting
the source terms for masses $m$ and $z$, respectively.
While the RMT partition function matches this only in the microscopic
limit\footnote{The quenched partition
  function has to be $m$- and $z$-independent for that, unlike the
  convention in \cite{AN}.}, 
it is tempting to assume that an RMT for $a^2_{6,7}\neq0$ can
be constructed at finite-$n$, which was recently achieved in
\cite{M11}. The integral transformed jpdf is non-trivial in general,
and special care has to be taken regarding the normalisation 
when computing spectral densities and
individual eigenvalue distributions of such RMT (see e.g. \cite{AV08}
for a similar discussion).

In particular it leads not
only to a shift of the masses, but partly of the eigenvalues which can
be viewed as extra auxiliary masses in the graded framework
\cite{ADSV} (c.f. \cite{KSV,M11}). If we define an RMT jpdf including
$a^2_{6,7}<0$ \cite{KSV} 
\bea
{\cal P}_{5}(\{x\};m;z;a_6;a_7) 
&=& \frac{1}{16\pi |a_6 a_7|}
\left(\int_{-\infty}^\infty\right)^2 dy_6
dy_7\ e^{-\frac{y_6^2}{16|a_6^2|}-\frac{y_7^2}{16|a_7^2|}}
{\cal P}_{5}(\{x+y_6\};m-y_6;z-y_7) ,\ \ \ \ \ \ \ \ 
\label{P5W67}
\eea
and a corresponding RMT partition function
we obtain for the $k$-point density
\bea
R_{k}(x_1,\ldots,x_k;m;z;a_6;a_7) 
&=& \frac{\left(\int_{-\infty}^\infty\right)^2 
dy_6 dy_7 \ e^{-\frac{y_6^2}{16|a_6^2|}-\frac{y_7^2}{16|a_7^2|}}
}{{\cal Z}_{5}(m;z;a_6;a_7)16\pi |a_6 a_7|}
{\cal Z}_{5}(m-y_6;z-y_7;a_{6}=0;a_7=0)
\nn\\
&&\times
R_{k}(x_1+y_7,\ldots,x_k+y_7;m-y_6;z-y_7;a_{6}=0;a_7=0)\ ,
\label{RkW67}
\eea
where we explicitly display all arguments here. We have
checked that the resulting microscopic density $k=1$ agrees with that
following from the resolvent in \cite{ADSV} in various cases.
Because eq. (\ref{EkRk}) is a liner map this translates as follows to
individual eigenvalues,
\bea
p_{k}(c,b;m;z;a_6;a_7) 
&=& \frac{\left(\int_{-\infty}^\infty\right)^2 dy_6 dy_7 
\ e^{-\frac{y_6^2}{16|a_6^2|}-\frac{y_7^2}{16|a_7^2|}}
}{{\cal Z}_{5}(m;z;a_6;a_7)16\pi |a_6 a_7|}
{\cal Z}_{5}(m-y_6;z-y_7;a_6=0;a_7=0)
\nn\\
&&\times
p_{k}(b+y_7,c+y_7;m-y_6;z-y_7;a_6=0;a_7=0)\ , 
\label{pkW67}
\eea
making our framework directly applicable in this setting too.


\subsection{The jpdf of $D_W$}\label{jpdfDW}
We now turn to the RMT for the non-Hermitian Wilson Dirac
operator following \cite{ADSV}, first at $W_{6,7}=0$:
\bea
{\cal Z}_W&\equiv&
\int dA\int dB \int dW \prod_{f=1}^{N_f}
\det[D_W +m_f\one_N]
\exp\left[-\ n\Tr WW^\dag-\ \frac{n}{2}\Tr(A^2+B^2)\right],
\label{ZW}\\
D_W &\equiv&
\left(\begin{array}{cc}
aA& W\\
-W^\dag&aB\\
\end{array}\right).
\label{DW}
\eea
The Hermitian random matrices $A=A^\dag$ and $B=B^\dag$ are of sizes
$n\times n$ and $(n+\nu)\times(n+\nu)$, respectively. Here the
parameter $a\geq0$ is not restricted and the masses $m_f$
can in principle be taken non-degenerate. The Dirac matrix
$D_W$ can only be 
quasi-diagonalised and we refer to \cite{KVZ} for details where the jpdf was
computed. Choosing $\nu\geq0$ without loss of generality, the $2n+\nu$
eigenvalues of $D_W$ consist of $n-l$ complex conjugate eigenvalue
pairs, of $l$ real right 
eigenvalues and of $l+\nu$ real left eigenvalues. Here $l$ can take any
value $0\leq l\leq n$, and we have to sum over all possible sectors. The
partition function can thus be written as follows \cite{KVZ} for $N_f=0$:
\bea
{\cal Z}_W &\equiv& \left(\int_{\mathbb C}\right)^{2n+\nu}
dz_{1R}\ldots dz_{nR}dz_{1L}\ldots dz_{n+\nu\,L} {\cal
  P}^{(N_f,\nu)}_W(z_{1R},\ldots, z_{nR},z_{1L},\ldots, z_{n+\nu\,L}) 
\label{ZWjpdf}\\
&=&\left(\int_{\mathbb C}\right)^{2n+\nu}
dz_{1R}\ldots dz_{nR}dz_{1L}\ldots dz_{n+\nu\,L}
\Delta_{2n+\nu}(\{z\})\nn\\
&&\times\det_{1\leq a\leq n,\,1\leq b\leq n+\nu,\,1\leq
  c\leq \nu}
\left[
\begin{array}{c}
g_{\mathbb C}(z_{aR},z_{bL})\delta^{(2)}(z_{aR}-z_{bL}^*) +
g_{\mathbb R}(x_{aR},x_{bL})\delta^{(1)}(y_{aR})\delta^{(1)}(y_{bL})\\
x^{c-1}_{bL}g_1(x_{bL})\delta^{(1)}(y_{bL})
\end{array}
\right]\ \ \ \ \ \nn\\
&&\nn\\
&\equiv& \sum_{l=0}^n \left(\int_{\mathbb C}\right)^{n-l}dz_1\ldots
dz_{n-l} \left(\int_{-\infty}^\infty\right)^{2l+\nu} dx_{1R}\ldots
dx_{lR}dx_{1L}\ldots dx_{l+\nu L}\nn\\ 
&&\times
{\cal P}^{(N_f,\nu)}_{W,l}(z_{1},\ldots, z_{n-l},x_{1R}\ldots,
x_{lR},z_{1}^*,\ldots, z_{n-l}^*,x_{1L},\ldots,x_{l+\nu\,L})\ . 
\label{PWldef}
\eea
Here we defined two different jpdf, with a fixed number of $l$ real
right eigenvalues ${\cal P}^{(N_f,\nu)}_{W,l}$, and without fixing
${\cal P}^{(N_f,\nu)}_{W}$. 
We note in passing that the latter jpdf in eq. (\ref{ZWjpdf}) can also
be written in terms of a single Pfaffian \cite{M11}. 
The one- and two-dimensional delta-functions in the variables
$z_{a}=x_{a}+iy_{a}$ inside the determinant assure that we sum over
all 
possible sectors. This sum is explicitly given in the last equation.
The respective weight functions are defined as \cite{KVZ}
\bea
g_{\mathbb C}(z_1,z_2)&\equiv&\sqrt{\frac{n^3}{4\pi a^2(1+a^2)}}
\frac{z_1^*-z_2^*}{|z_1-z_2|}  \exp\left[-\frac{n(x_1+x_2)^2}{4a^2}
-\frac{n(y_1-y_2)^2}{4} \right],\\
g_{\mathbb R}(x_1,x_2)&\equiv&\sqrt{\frac{n^3}{4\pi a^2(1+a^2)}}
\frac{z_1^*-z_2^*}{|z_1-z_2|}\frac12   \exp\left[-\frac{n(x_1+x_2)^2}{4a^2}
-\frac{n(x_1-x_2)^2}{4} \right],\nn\\
&&\times \erfc\left[\sqrt{n(1+a^2)}\frac{|x_1-x_2|}{2a}\right],\\
g_1(x)&\equiv& \sqrt{\frac{n}{2\pi a^2}}\exp\left[-\frac{nx^2}{2a^2}\right].
\label{gdefs}
\eea 
The fact that we have to distinguish between left and right
eigenvalues of $D_W$ leads to more possibilities of defining density
correlation functions, compared to eq. (\ref{Rk}). 
When defining the density by inserting a delta-function into the jpdf
we encounter a density of complex, of left and of right
real eigenvalues, respectively. 
For example following \cite{KVZ} the density of complex eigenvalues,
$R_{1\mathbb C}$, and the 
density of real right eigenvalues $R_{1{\mathbb R}R}$ is defined as
\bea
R_{1{\mathbb R}R}(x)\delta^{(1)}(y) +\frac12 R_{1\mathbb C}(z)
&\equiv&
\left(\int_{\mathbb C}\right)^{2n+\nu} dz_{1R}\ldots dz_{n+\nu\,L}
     {\cal P}_{W}^{(N_f,\nu)}(\{z\}) 
\delta^{(2)}(z-z_{1R})
\ ,
\label{R1rdef}
\eea
and likewise for the density of real left eigenvalues $R_{1{\mathbb
    R}L}$. It is clear by comparing to eq. (\ref{PWldef}) that in
principle both densities get contributions from all sectors $l\geq0$.
Note that in \cite{KVZ} a particular combination of these densities
was defined, the so-called density of chirality: 
\be
R_{{\mathbb R}\chi}(x)\equiv R_{1{\mathbb R}L}(x)-R_{1{\mathbb R}R}(x)\ .
\label{Rchidef}
\ee
Higher $k$-point correlation functions have to be split as well. For
example the 2-point function will have 3 contributions: to find 2
complex eigenvalues, one complex and one real, and two real
eigenvalues; and so on for higher $k$.

We would like to point out that if one is only interested in the
correlation functions of real eigenvalues as we are here, 
at least approximately one
can obtain the situation of having a jpdf of only finitely many $\nu$
real left eigenvalues in the large-$n$ limit. Hence for this setting
expanding the Fredholm determinant to obtain the first individual
eigenvalue truncates at exactly $\nu$ terms. 
In \cite{KVZ} the probability to obtain $N_{add}$ more real
(right and left) eigenvalues in
addition to $\nu$ was computed as a function of the rescaled spacing
$\ha=a\sqrt{n/2}$ and of $\nu$, see figure 4 in \cite{KVZ}. For example for a
value of $\ha=0.1$ there are less than $2\%$ additional real eigenvalues
for $\nu=0$ (and even less for $\nu>0$), that is $2\%$ contributions
from sectors with $l>0$. 
Restricting ourselves to the sector $l=0$ we can thus define an
approximate jpdf for the real eigenvalues valid for small $\ha$: 
\be
{\cal P}^{(N_f,\nu)}_{W,{\mathbb R}}(x_{1L},\ldots,x_{\nu\,L})\ \equiv\
\left(\int_{\mathbb C}\right)^{n}dz_1\ldots dz_{n}
{\cal P}^{(N_f,\nu)}_{W,l=0}(z_{1},\ldots, z_{n},z_{1}^*,\ldots,
z_{n}^*,x_{1L},\ldots,x_{\nu\,L}) \ .
\ee
We can now follow eqs. (\ref{Rk}) and (\ref{EkRk}) with $n=\nu$
there, and we only need to insert the $k$-point densities of real left
eigenvalues for up to $k=\nu$, with $\nu$ being small and finite. 
Note that in this approximation we have that
\be
R_{{\mathbb R}\chi}(x)\approx R_{1{\mathbb R}L}(x) \gg R_{1{\mathbb R}R}(x)\ .
\label{RX=RL}
\ee
This is consistent when comparing to figures 5 and 6 in ref \cite{KVZ}
for small values of $\ha$. 

Regarding the introduction of non-zero values for $W_{6,7}$ the same
method discussed at the end of the previous subsection applies 
after adding a $\gamma_5$-mass term in eq. (\ref{DW}),
and we refer to \cite{KSV,M11} for more details on the
transformed densities.


\sect{Examples for individual real eigenvalues 
in the microscopic limit
}\label{examples}

In this section we will give explicit examples for individual
eigenvalue distributions. Because RMT is only equivalent to
Wilson XPT in the epsilon regime 
after
taking the microscopic large-$n$
limit, we directly give results in this limit and refer to the
literature for details. 
We present 
only quenched higher $k$-point
density correlation functions as an example,
which we need in
our expansion. 
However, as it should have become clear from the previous
sections our method is not restricted to this case, 
it could immediately be applied to unquenched densities too.
As a further
simplification we choose to work here with $W_6=W_7=0$. Again this
restriction is not a principle one, and we have shown how to lift it
following \cite{ADSV}. To keep everything as simple and transparent as
possible we choose to work with $W_8\neq0$ only. 

The examples for the 
quenched
Hermitian Wilson Dirac operator are presented in
subsection \ref{ex5}, and the examples for the distribution of real
eigenvalues of the Wilson Dirac operator follow in subsection
\ref{exW}.

\subsection{Hermitian Wilson Dirac operator $D_5$}\label{ex5}

The computation of the microscopic large-$n$ limit was discussed in
great detail in ref. \cite{AN} which we shall not repeat here,
recalling only the results. The only difference to ref. \cite{AN} here
is the extra factor of $2n$ in the exponent of the weight function in
eq. (\ref{Z5}). We have chosen this convention here, to be achieved by
a simple rescaling of all matrix elements, in order to have the same
microscopic rescaling of the eigenvalues $x$, masses $m$ and lattice
spacing $a$ as for $D_W$ in the conventions of ref. \cite{KVZ}. 

The microscopic quantities denoted by hat are defined as follows:
\be
\hat{x}\equiv{2n}\,x\ ,\ \ \hat{m}\equiv{2n}\,m\ ,\ \
\hat{a}\equiv \frac12\sqrt{2n}\,a \ .
\label{micro}
\ee
The map to the LEC in the chiral Lagrangian is provided by
\be
\hat{x}\equiv xV\Sigma\ ,\ \ \hat{m}\equiv mV\Sigma\ ,\ \
\hat{a}^2\equiv a^2_{lat}V W_8 \ ,
\label{micro2}
\ee
where $V$ is the volume, $\Sigma$ the chiral condensate, $a_{lat}$ the
lattice spacing and $W_8$ the LEC.
The microscopic $k$-point density correlation functions have to be
rescaled appropriately with $n$, 
\be
\rho_{S,k}(\hx_1,\ldots,\hx_k)\equiv\lim_{n\to\infty}\frac{1}{(2n)^k}R_k(x_1,
\ldots, x_k)\ , 
\label{rhokdef}
\ee
with all the arguments on the right hand side rescaled according to
eq. (\ref{micro}). The rescaled microscopic gap probabilities are
defined as 
\be
E_{S,k}(\hat{b},\hat{c})\equiv\lim_{n\to\infty}E_k(b,c)\ ,
\label{Ekdef}
\ee
and correspondingly the resulting rescaled individual eigenvalue
distributions $p_{S,k}$ and $q_{S,k}$ follow. 
Hence the boundaries of the interval $[b,c]$ considered have to be
rescaled in the same way as the eigenvalues in eq. (\ref{micro}). 

We will now give the building blocks of the limiting microscopic
kernel. All 3 kernels in eq. (\ref{3kernels}) can be expressed through
the kernel $D_n(x,y)$, the weight function $F(x)$ eq. (\ref{Fdef}) and
integrals thereof, plus some extra terms for $\nu=1$. 
Referring to \cite{AN} for details we have for our first example at $\nu=0$:
\bea
D^{\nu=0}_S(\hz,\hy)&=&\frac{1}{32\sqrt{2\ha^2\pi}}
\exp\left[\frac{-\hz^2-\hy^2}{16\ha^2}\right] 
\left(\int_{-\infty}^\infty\right)^2 
\frac{ds\,dr}{{\pi}}\e^{-s^2-r^2}(\hy-\hz+4i\ha(s-r))\nn\\
&&\times
\int_0^1dt
I_0\Big( \sqrt{t(\hat{m}^2-(\hat{z}+4ir\hat{a})^2)}\Big)
I_0\Big( \sqrt{t(\hat{m}^2-(\hat{x}+4is\hat{a})^2)}\Big)\ ,
\label{Dmic}\\
S^{\nu=0}_S(\hx,\hy)&=&
\int_{-\infty}^\infty
d\hat{z}\,{\exp\left[\frac{(\hx-\hz)^2}{32\ha^2}\right]}
\left(\erf\Big(\frac{\hx-\hz+2\hm}{4\sqrt{2\ha^2}}\Big)
+\erf\Big(\frac{\hx-\hz-2\hm}{4\sqrt{2\ha^2}}\Big)
\right)D^{\nu=0}_S(\hz,\hy),\ \ 
\label{Smic}\\
I^{\nu=0}_S(\hx,\hy)&=& -\int_{-\infty}^\infty d\hz
F_S(\hy-\hz)S^{\nu=0}_S(\hx,\hz) 
\ -\ F_S(\hx-\hy)\ ,
\label{Imic}
\eea
where
\be
F_S(\hx)
=\exp\Big[\frac{\hx^2}{32\ha^2}\Big]\left[
\erf\left(\frac{\hx}{4\sqrt{2\ha^2}}+\frac{\hm}{2\sqrt{2\ha^2}}\right)
+\erf\left(\frac{\hx}{4\sqrt{2\ha^2}}-\frac{\hm}{2\sqrt{2\ha^2}}\right)\right]
\label{Fmic}
\ee
is the rescaled, microscopic weight eq. (\ref{Fdef}). One of the
integrals can be computed analytically, 
\be
\int_0^1 dt
I_0(X\sqrt{t})I_0(Y\sqrt{t})=\frac{2XI_0(Y)I_1(X)-2YI_0(X)I_1(Y)}{X^2-Y^2}\ .
\ee
However, this is not always advantageous when plotting the
densities. From these building blocks we obtain the following
expressions for the densities:
\bea
\rho_{S,1}(\hx)&=&S_{S}(\hx,\hx)\ ,
\nn\\
\rho_{S,2}(\hx,\hy)&=&S_S(\hx,\hx)S_S(\hy,\hy)+I_S(\hx,\hy)D_S(\hx,\hy)-
S_S(\hx,\hy)S_S(\hy,\hx)\ ,
\label{rhoS1,2}
\eea
and for general $k$
\be
\rho_{S,k}(\hx_1,\ldots,\hx_k)= \Pf_{1\leq i,j\leq k}
\left[
\begin{array}{cc}
I_S(\hx_i,\hx_j)& S_S(\hx_i,\hx_j)\\
-S_S(\hx_j,\hx_i)& -D_S(\hx_i,\hx_j)\\
\end{array}
\right].
\label{rhoSPf}
\ee

\begin{figure}[-h]
\centerline{\epsfig{figure=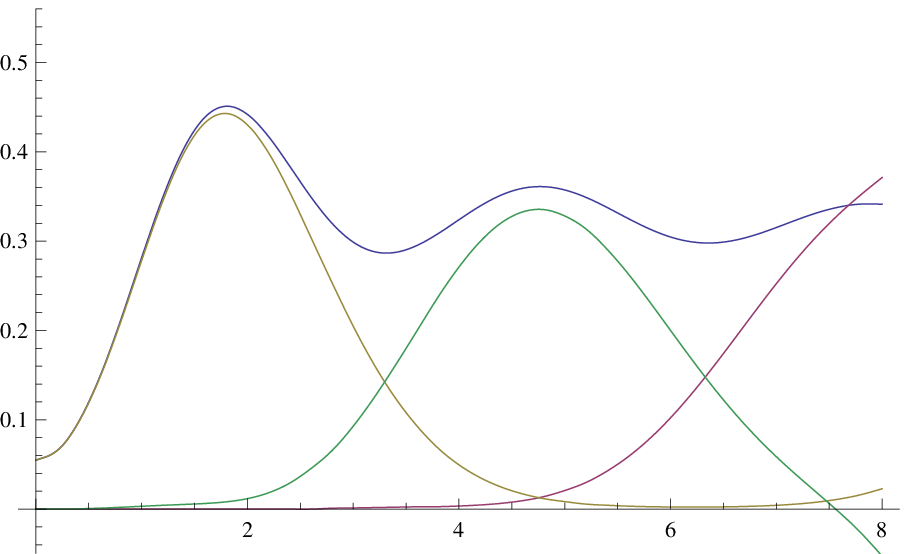,width=20pc}
\put(-230,150){$\rho_{S,1},\,p_{S,k=1,2,3}$}
\put(10,10){$\hx$}
}
\centerline{\epsfig{figure=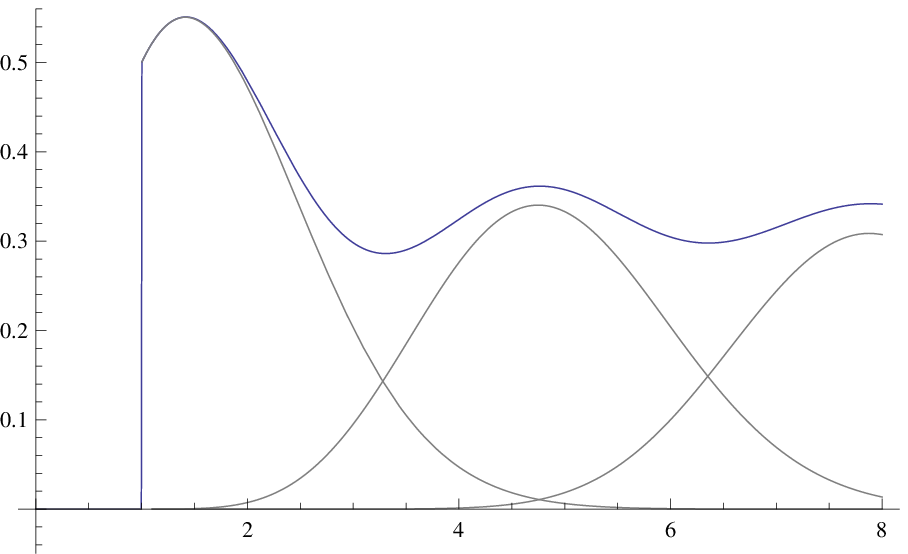,width=20pc}
\put(-230,150){$\rho_{S,1},\,p_{S,k=1,2,3}|_{\ha=0}$}
\put(10,10){$\hx$}
}
\caption{Top plot: example for the approximated first, second and third positive
  $D_5$ eigenvalue: $p_{S,1}(0,\hx)$ (left yellow curve), $p_{S,2}(0,\hx)$
  (middle green curve) and $p_{S,3}(0,\hx)$ (right pink curve),
from eqs. (\ref{pS1})-(\ref{pS3}) respectively, in the quenched theory
with $\nu=0$. The parameter values chosen are $\hm=1$ and
$\ha=0.25$. We also show the spectral density $\rho_{S,1}(\hx)$ (top
blue curve) from eq. (\ref{rhoS1,2}) and (\ref{Smic}) for
comparison, which is symmetric around the origin.
Bottom plot:
the spectral density eq. (\ref{rhonua0}) and exact
individual eigenvalue distributions (eq. (\ref{p1nu0a0}), \cite{DN}) 
for $D_5$ at
$\ha=0$, shifted according to eq. (\ref{rhoshift}).}
\label{p123rho3nu0}
\end{figure}
As our first example we will now give the distribution of the first,
second and third positive eigenvalue of $D_5$ for $\nu=0$, using only
the first 3 densities $\rho^{\nu=0}_{S,j}$ with $j=1,2,3$ as an
approximation when truncating the Fredholm Pfaffian expansion.
In order to 
describe
the first positive eigenvalue from the origin we
choose $[0,c]$ as the interval for the gap probability, differentiate
with respect to the upper bound $c$ and rescale $2nc=\hx$. Using
eq. (\ref{pkEk}) at this level of truncation we obtain on ${\mathbb
  R}_+$
\bea
p_{S,1}(0,\hx)&\approx& \rho_{S,1}(\hx)-\int_0^{\hx} d\hy \rho_{S,2}(\hx,\hy)
+\frac12 \int_0^{\hx} \int_0^{\hx} d\hy d\hz \rho_{S,3}(\hx,\hy,\hz)\ ,
\label{pS1}\\
p_{S,2}(0,\hx)&\approx& \int_0^{\hx} d\hy \rho_{S,2}(\hx,\hy)
-\int_0^{\hx} \int_0^{\hx} d\hy d\hz \rho_{S,3}(\hx,\hy,\hz)\ ,
\label{pS2}\\
p_{S,3}(0,\hx)&\approx&\frac12
\int_0^{\hx} \int_0^{\hx} d\hy d\hz \rho_{S,3}(\hx,\hy,\hz)\ .
\label{pS3}
\eea
Their generating cumulative distributions are obvious.
Note that when summing up these first 3 approximate individual eigenvalue
distributions eqs. (\ref{pS1}) - (\ref{pS3}) we trivially get the
exact density. 
Would we add the exact
individual eigenvalue distributions instead, this identity would only hold
after summing up all eigenvalues, $\rho_{S,1}(\hx)=\sum_{j=0}^\infty
p_{S,j}(\hx)$.

Inserting the above expressions eqs. (\ref{rhoS1,2}) and
(\ref{rhoSPf}) together with eqs. (\ref{Dmic})-(\ref{Fmic}) leads to
the curves shown in figure \ref{p123rho3nu0} (top plot).
From it we see that our approximation gives well localised
curves for the 1st and 2nd eigenvalue. While the 1st eigenvalue
smoothly touches zero at around $\hx\approx6$, clearly the approximation gets
worse for the 2nd and even more so for the 3rd eigenvalue. To get a
similar quality for the 2nd eigenvalue we would need to include up to
terms containing $\rho_{S,4}$, and for the 3rd eigenvalue terms up to
$\rho_{S,5}$. The fact that the left tail of the first positive
eigenvalue is not shown for this value of $\ha$
is an artefact of counting eigenvalues from the
origin. If we were to use another interval (say $[-2,\hx]$) we could
also determine this tail. However, the origin is a good
point to choose since in the limit $\ha\to0$ chiral symmetry is restored.

\begin{figure}[-h]
\centerline{\epsfig{figure=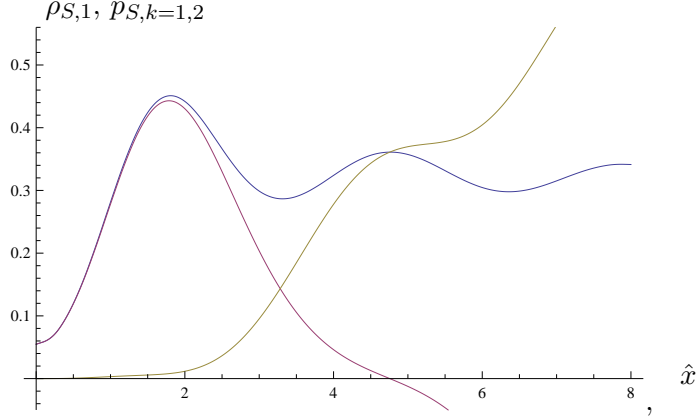,width=20pc},
\put(-230,150){$\rho_{S,1},\,p_{S,k=1,2}$}
\put(10,10){$\hx$}
}
\caption{For comparison we show the same plot as in the top
  fig. \ref{p123rho3nu0} for the 1st and 2nd eigenvalue, using only an
  approximation up the second term in eqs. (\ref{pS1}) - (\ref{pS2})
  with $\rho_{S,1}$ and $\rho_{S,2}$ at otherwise equal parameter
  values. It can be nicely seen how the quality of the approximation
  improves systematically from this plot to the previous plot, moving
  the breakdown point from $\hx_c\approx4.5$ here to
  $\hx_c\approx7$ there.
}
\label{p123rho2nu0}
\end{figure}

\begin{figure}[-h]
\centerline{
\epsfig{figure=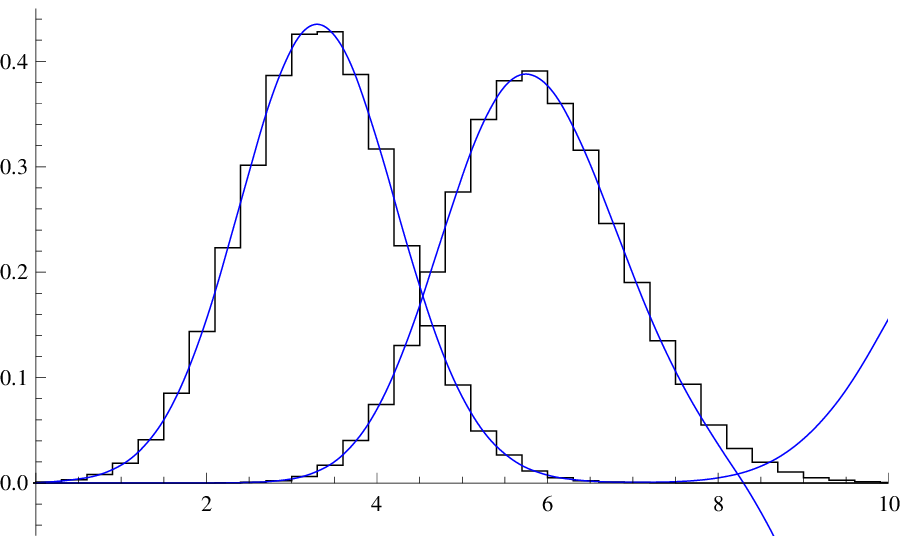,width=20pc}
\epsfig{figure=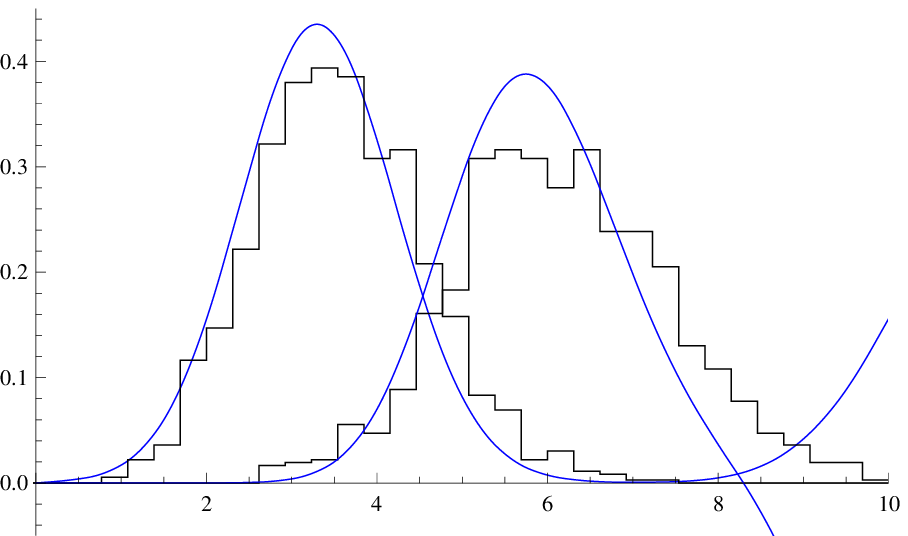,
width=20pc}
\put(-480,145){$p_{S,k=1,2}$}
\put(-230,145){$p_{S,k=1,2}$}
\put(-250,-10){$\hx$}
\put(-5,-10){$\hx$}
} 
\caption{Left plot: Numerically generated first and second RMT
  eigenvalue distributions (histograms) with $10^6$ matrices of size $n=100$
from \cite{DHS}  vs. our analytical curves.
Right plot: comparison of lattice data (histograms) from \cite{DHS} with our
analytical RMT curves. The data set is for volume $20^4$,
$\beta_{I_w}=2.79$, where the predicted values for $\Sigma=216$, $\hm=3.5$,
$\ha=0.35$ 
determined from another volume are used, and we refer to \cite{DHS} for
a more detailed discussion.}
\label{numRMT+latt}
\end{figure}

It is clearly seen when the approximation breaks down at latest: that is when
$p_{S,1}$ increases again from zero for large $\hx$,  when $p_{S,2}$
becomes negative for the second eigenvalue, and when $p_{S,3}$ exceeds
the spectral density which is an upper bound to all individual
eigenvalue distributions. This breakdown happens roughly at the same
point $\hx_c\approx 7$, up to which we can trust our approximated
individual distributions (see also fig. \ref{numRMT+latt} left).
This is because from this point onwards the 4th eigenvalue would start
contributing which is zero to our order of approximation.
Nevertheless even the first part of the ascending curve of the 3rd
eigenvalue could be used for fitting purposes to this order.
In order to illustrate the described convergence of our 
approximation scheme
we show in fig. \ref{p123rho2nu0} 
the same plot as
fig. \ref{p123rho3nu0} 
(top plot), using one term less in the approximation:
the density and the two-point function.

Let us also compare our approximate distributions in
fig. \ref{p123rho3nu0} (top plot) with the corresponding first, second and third
eigenvalues when the lattice spacing $\ha=0$, as shown in
fig. \ref{p123rho3nu0} (bottom plot). In this limit also the individual
eigenvalue distributions and microscopic density are known exactly:
\bea
\tilde{p}^{\nu=0}_{S,1}(\hx)\Big|_{\ha=0}&=&\frac{\hx}{2} \exp\left[
  -\frac{\hx^2}{4}\right]\ ,
\label{p1nu0a0}\\
\tilde{p}^{\nu=1}_{S,1}(\hx)\Big|_{\ha=0}&=&\frac{\hx}{2} \exp\left[
  -\frac{\hx^2}{4}\right] I_2(\hx)\ ,
\label{p1nu1a0}\\
\tilde{\rho}_{S,1}(\hx)\Big|_{\ha=0}&=& \frac{\hx}{2}\left(
J_\nu(\hx)^2-J_{\nu-1}(\hx)J_{\nu+1}(\hx)\right)\ .
\label{rhonua0}
\eea
Here we only give the first eigenvalue distributions following
\cite{DNW}. The expressions for higher eigenvalues can be found in
\cite{DN}. In order to compare to the $D_5$ eigenvalues which contain
the mass, cf. eq. (\ref{D5}), we need to shift the density
$\tilde{\rho}_{S,1}$  eq. (\ref{rhonua0}) as in \cite{ADSV},
\be
{\rho}_{S,1}(\hx)|_{\ha=0}\ =\  
\frac{|\hx|}{\sqrt{\hx^2-\hm^2}}\tilde{\rho}_{S,1}
\left(\sqrt{\hx^2-\hm^2}\right)\Big|_{\ha=0}\Theta(|\hx|-\hm)\ ,
\label{rhoshift}
\ee
and the individual eigenvalue distributions accordingly. These
shifted distributions are plotted in fig. \ref{p123rho3nu0} (bottom
plot). It can be clearly seen that the first eigenvalue is most sensitive to
the influence of $\ha\neq0$, for which we have the best
approximation.

We would like to compare our new analytical predictions for individual
$D_5$ eigenvalues with RMT and lattice data, following very recent comparisons
\cite{DHS,DWW} where numerically generated curves for RMT were used.
As an illustration we use the RMT and lattice
data from fig. 8 in ref. \cite{DHS}, shown
in our figure \ref{numRMT+latt} left and right, respectively. 
Our approximate analytical curves for
the individual eigenvalues match very well the numerically generated
RMT histograms from  \cite{DHS} in the left plot as they should.
In the comparison to the lattice data - these are predictions from
other data in \cite{DHS} and not
fits to these curves - one can see a slight systematic shift 
to the left.
Our analytic curves clearly are a very sensitive tool to fit lattice
data. Compared to fits of the spectral density also used in
\cite{DHS,DWW}, individual
eigenvalues have an unambiguous normalisation to unity.

\begin{figure}[-h]
\centerline{\epsfig{figure=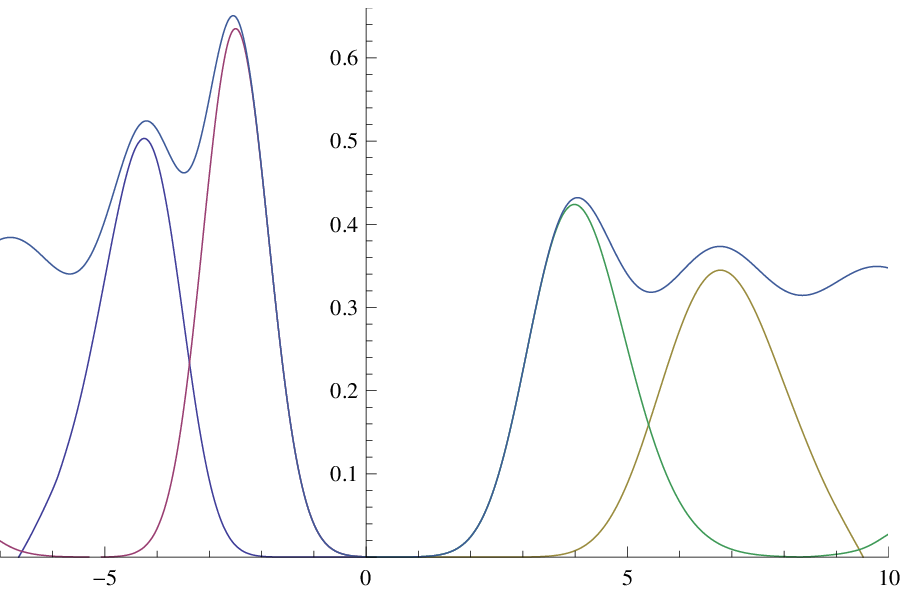,width=20pc}
\put(-150,160){$\rho_{S,1},\,p_{S,k=1,2},\,q_{S,k=1,2}$}
\put(10,5){$\hx$}
}
\centerline{\epsfig{figure=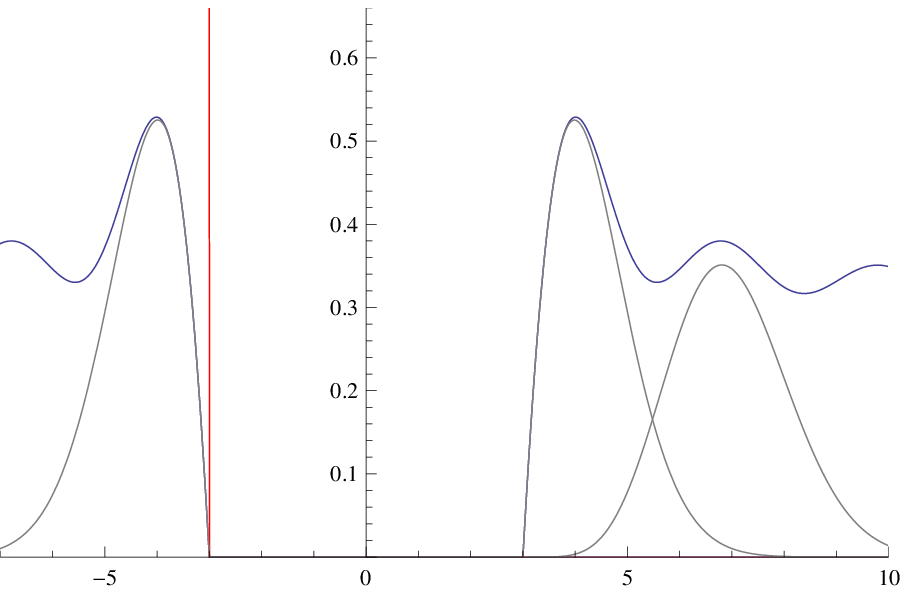,width=20pc}
\put(-150,160){$\rho_{S,1},\,p_{S,k=1,2},\,q_{S,k=1,2}|_{\ha=0}$}
\put(10,5){$\hx$}
}
\caption{Top plot: the spectral density vs. the first two positive 
  and negative individual eigenvalues $p_{S,k=1,2}(0,\hx)$ and 
$q_{S,k=1,2}(\hx,0)$ respectively, 
for $\hm=3$ and $\ha=0.25$ at $\nu=1$. Negative values for the second
eigenvalues are clipped.
Bottom plot: the same curves at $\ha=0$, with the spectral density
and exact first and second eigenvalue given (eqs. (\ref{rhonua0}),
(\ref{p1nu1a0}), \cite{DN}), respectively. 
The location of the delta-function due
to the exact zero eigenvalues shifted to  $-\hm$ is indicated by the
vertical line.
}
\label{p12rho3nu1}
\end{figure}

Let us now turn to our second example at $\nu=1$. Here the building
blocks for the kernel read
\cite{AN}:
\bea
D^{\nu=1}_S(\hx,\hy)&=&
D^{\nu=0}_S(\hx,\hy)+\frac{1}{16\ha\pi\sqrt{2\pi}}
\exp\left[-\frac{\hx^2+\hy^2}{16\ha^2}\right]
\left\{ \int_{-\infty}^\infty ds
\e^{-s^2}I_0\Big( \sqrt{\hat{m}^2-(\hy+4is\ha)^2}\Big)
\right.\nn\\
&&\times\left. \int_{-\infty}^\infty dr\ \e^{-r^2}
\frac{\hm+\hx+4ir\ha}{\sqrt{\hm^2-(\hx+4ir\ha)^2}}
I_1\Big( \sqrt{\hat{m}^2-(\hx+4ir\ha)^2}\Big)
\ -\ (\hx\leftrightarrow\hy)\right\},
\label{DSnu1}\\
S_S^{\nu=1}(\hx,\hy)&=&\int_{-\infty}^\infty d\hz 
F_S(\hx-\hz)D^{\nu=1}_S(\hz,\hy)
+\frac{1}{4\pi\ha}\e^{-\frac{\hy^2+2\hx\hm+\hm^2}{16\ha^2}}
\int_{-\infty}^\infty ds \e^{-s^2}I_0\Big(
\sqrt{\hat{m}^2-(\hy+4is\ha)^2}\Big),
\nn\\
&&\label{SSnu1}\\
I_S^{\nu=1}(\hx,\hy)&=&
- \int_{-\infty}^\infty d\hz F_S(\hy-\hz)S^{\nu=1}_S(\hx,\hz)\ -\ F_S(\hx-\hy)\
+\frac{1}{4\pi\ha}\int_{-\infty}^\infty d\hz
\exp\left[-\frac{\hz^2+\hm^2}{16\ha^2}\right]
\nn\\
&&\times \left(F_S(\hx-\hz)\e^{-\frac{\hy\hm}{8\ha^2}}
-F_S(\hy-\hz)\e^{-\frac{\hx\hm}{8\ha^2}}\right)
\int_{-\infty}^\infty ds \e^{-s^2}I_0\Big( \sqrt{\hat{m}^2-(\hz+4is\ha)^2}\Big).
\label{ISnu1}
\eea
We simply have to insert these into eq. (\ref{rhoSPf}) to obtain the
densities, and into eqs. (\ref{pS1}) - (\ref{pS3}) to obtain the
individual eigenvalues to our order of approximation.
The corresponding curves are shown in figure \ref{p12rho3nu1} (top plot).
Note that the spectral density is no longer symmetric with respect to
the origin.
This is due to the broadening by $\ha>0$ of the single zero eigenvalue ($\nu=1$)
which is shifted to the location of the mass, $\hx=-\hm$. The first
and second negative eigenvalue is determined separately using the quantities
eq. (\ref{qk}), that is by differentiating the
gap probability with respect to the lower bound of the gap $[b,0]$.
When comparing to the same curves at $\ha=0$ in figure
\ref{p12rho3nu1} (bottom plot), the first (``non-zero'') negative eigenvalue
there becomes the second negative eigenvalue for $\ha>0$ in the top
plot, and receives considerable corrections as well.

\subsection{Real eigenvalues of the Wilson Dirac operator $D_W$}\label{exW}

For $k$-point correlation functions of left (or right) real
eigenvalues of $D_W$ 
much less was known compared to the
previous subsection for $D_5$. Explicit results 
included only
the quenched
$k=1$ and $(k=2)$-point function for arbitrary $\nu$. More general
results for all quenched $k$-point functions have been announced
\cite{KVZ} and appeared now \cite{M11}.
For $k=1$ it is either given as the discontinuity of the corresponding
resolvent \cite{ADSV}, or more explicitly as a double integral in
\cite{KVZ}. For $k=2$ its resolvent is given as a 4-fold integral over
a $4\times4$ determinant times some rational function \cite{SV2011}. 
As
expected this result can be factorised 
and brought into a Pfaffian
form \cite{M11}, 
in analogy to \cite{M}. Plotting the integral in the form of \cite{SV2011}
would still be a formidable task. It would give us a good approximation to
the first real $D_W$ eigenvalue and to the ascent of the second one.

Rather than awaiting more tractable results announced
in \cite{KVZ} 
that have appeared after finishing this paper \cite{M11},
let us illustrate our method in the small
$\ha$-regime. Here it was shown \cite{ADSV} that the microscopic limit of the
spectral density $R_{{\mathbb R}\chi}(x)$ becomes a $\nu\times\nu$ GUE
distribution (just as the broadening of the zero-modes of $D_5$). 
We therefore expect that the jpdf will be given by that
of the GUE for $\nu\times \nu$ matrices. This is based on the analogy
to $D_5$, where the factorisation
of the jpdf into a GUE part for the broadened zero-modes and a chiral GUE
part for the remaining eigenvalues was shown in \cite{AN}. We thus
can write approximately for $l=0$ and $\nu\geq1$
\be
{\cal P}_{W,{\mathbb R}}(x_{1L},\ldots,x_{\nu\,L})\approx
\det_{1\leq i,j\leq \nu}\left[K_{\nu}(x_{iL},x_{jL})\right]\ ,
\label{GUEjpdf}
\ee
where the GUE kernel inside the determinant is given by
\be
K_{\nu}(x,y)\ \equiv\ \frac{1}{4\ha\sqrt{\pi}}
\exp\left(-\frac{1}{32\ha^2}(x^2+y^2)\right)
\sum_{j=0}^{\nu-1}\frac{1}{2^j j!}H_j\left(\frac{x}{4\ha}\right)
H_j\left(\frac{y}{4\ha}\right)\ .
\label{GUEkernel}
\ee
In this regime all microscopic eigenvalues are measured in units of
$4\ha$. The spectral density
\be
\rho_{1{\mathbb R}L}(x) = K_{\nu}(x,x)\ ,
\ee
is then simply the Wigner semi-circle for finite $n=\nu$. Higher
$k$-point density correlation functions follow easily from
eq. (\ref{GUEjpdf}),
\be
\rho_{k{\mathbb R}L}(x_{1\,L},\ldots,x_{k\,L})\ =\
\det_{1\leq i,j\leq k}\left[ K_{\nu}(x_{iL},x_{jL})\right]\ ,
\ee
for $k=1,\ldots,\nu$.
Based on these expressions, we can now give all individual eigenvalue
distributions 
by a finite number of terms for the gap
probability. In this case it is a Fredholm determinant of size
$\nu\times\nu$, which after expansion gives rise to $\nu+1$ terms.

\begin{figure}[-h]
\centerline{\epsfig{figure=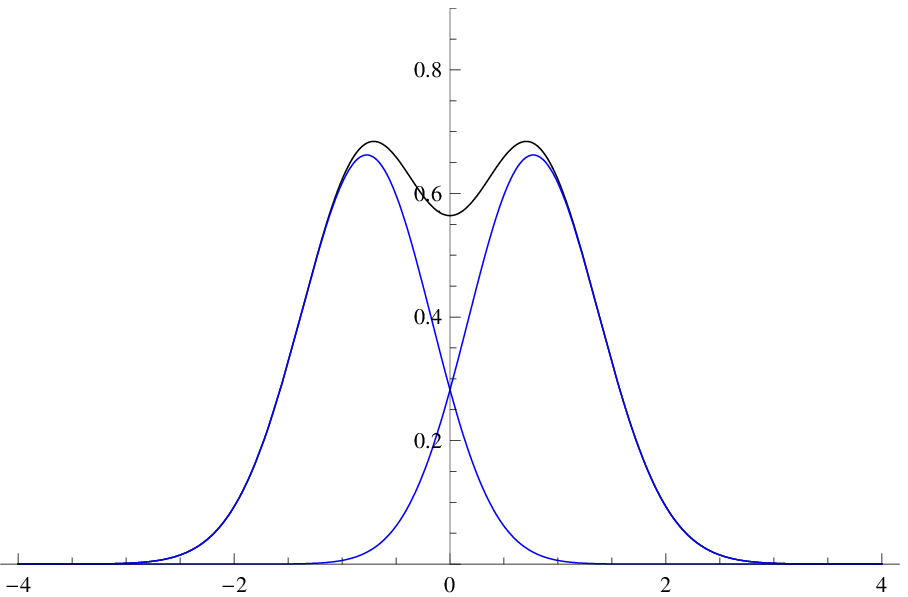,width=20pc}
\epsfig{figure=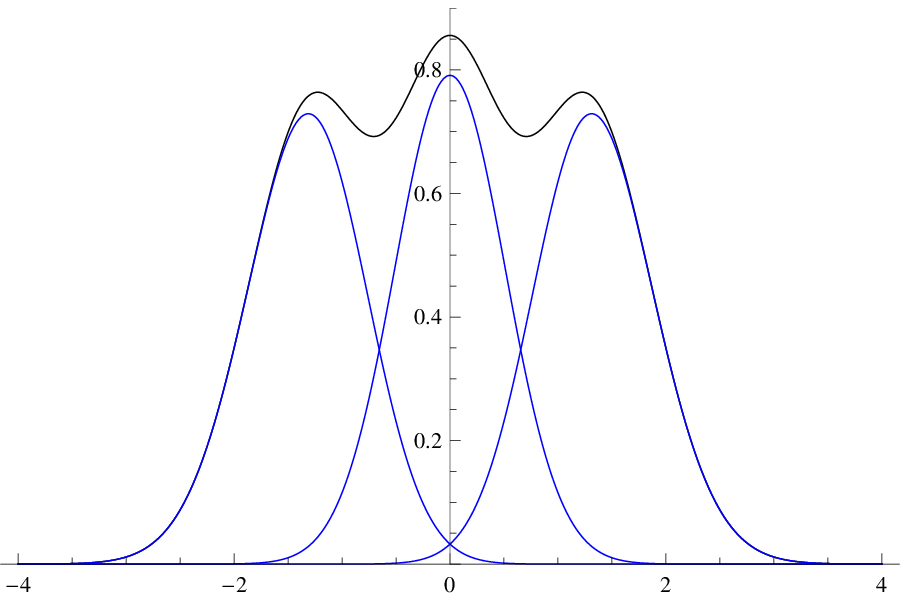,width=20pc}
\put(-370,160){$\rho_{1{\mathbb R}L},\,p_{S,k=1,2}$}
\put(-120,160){$\rho_{1{\mathbb R}L},\,p_{S,k=1,2,3}$}
\put(-250,-10){$\hx$}
\put(-5,-10){$\hx$}
}
\caption{The approximated density of real $D_W$ eigenvalues for $\nu=2$ (left
  plot) and $\nu=3$ (right plot). The spectral density is given by the
  finite-$\nu$ GUE semi-circle, with the individual eigenvalues
  obtained as described in the text. All quantities are measured in
  units $4\ha$. 
.}
\label{rhopGUE}
\end{figure}

In figure \ref{rhopGUE} we show the spectral density and the
individual eigenvalues for $\nu=2,3$.
As the gap interval we have to chose here $(-\infty,c]$.
For $\nu=2$ for example the distribution of the first and second
eigenvalues given in eqs. (\ref{p1example}) and (\ref{p2example}) are
exact, without the dots. For $\nu=3$ we can use eqs.  (\ref{pS1}) -
(\ref{pS3}) for the first three eigenvalues which are now exact, and
so on for higher $\nu$.

We mention in passing that for a jpdf with such a small and fixed
number of eigenvalues 
there exists an alternative computation instead of the Fredholm
determinant. The individual eigenvalue distributions could also be
computed as integrals over the ordered set of eigenvalues,
\bea
p_{1}(x_1) &\sim& \int_{x_1}^{\infty} dx_2 \int_{x_2}^{\infty} dx_3
\ldots \int_{x_{\nu-1}}^{\infty} dx_\nu
{\cal P}_{W,{\mathbb R}}(x_{1},\ldots,x_{\nu})\Big|_{x_1<\ldots<x_\nu}\ ,\\
p_{2}(x_2) &\sim& \int_{-\infty}^{x_2} dx_1 \int_{x_2}^{\infty} dx_3
\ldots \int_{x_{\nu-1}}^{\infty} dx_\nu
{\cal P}_{W,{\mathbb R}}(x_{1},\ldots,x_{\nu})\Big|_{x_1<\ldots<x_\nu}\ ,
\label{pkalt}
\eea
etc., and then normalising them appropriately.

We note that the computation of the individual eigenvalue
distributions based on eq. (\ref{GUEjpdf}) also applies to the $\nu$
broadened zero-eigenvalues of $D_5$, due to the approximate density
to $R_{{\mathbb R}\chi}(x)\approx R_{5}(x)$ for small $\ha$ close to
$\hx=-\hm$ \cite{ADSV}. One simply has to shift the argument of
the Hermite polynomials in eq. (\ref{GUEkernel}) as $\hx\to \hx+\hm$.

\sect{Conclusions}\label{conc}

We have set up a framework to analytically compute the
distribution of individual real  eigenvalues, both for
the Wilson Dirac operator $D_W$ and its Hermitian counterpart
$D_5$ in the epsilon regime of Wilson chiral 
perturbation theory. 
These distributions are given in a perturbative expansion of
$k$-point density correlation functions and integrals thereof. For
small lattice spacing and for the real $D_W$ eigenvalues this expansion
truncates after $\nu+1$ terms where $\nu$ plays the r\^ole of chirality on
the lattice. 
As an example for quenched $D_5$ 
all $k$-point density correlation functions are 
presented for
$\nu=0,1$ 
and we have given the resulting distributions of the
first two positive and two negative eigenvalues as a function of the low
energy constant $W_8$ times the lattice spacing squared
as an illustration. Our results very well describe previous numerical
results in the literature to the given order. 
We have 
briefly
discussed how to include the
effects of $W_{6,7}$ in our framework. Hopefully this setup will become
useful in the precise determination of all low energy constants to
leading order. 
This is especially so 
because more analytic density correlation functions have just become available
after finishing this paper \cite{M11}.

It would be very interesting to try to understand the mismatch
discussed in \cite{DHS} 
between lattice data and the analytical distribution of a single
real eigenvalue of $D_W$
(which is equal to the density $R_{{\mathbb R}\chi}$ as in that case
$\nu=1$). However, considering possible 
higher order effects in chiral perturbation theory is beyond the scope
of this article.

\vspace{0.5cm}
{\sc Acknowledgements}:~
The participants of the workshop
{``Chiral dynamics with Wilson fermions''} at ECT* Trento in October
2011 are thanked
for inspiring talks and discussions. We would also like to thank the
authors of \cite{DHS} for making some of their numerical lattice
and RMT data available to us.
We are indebted to Mario Kieburg, Kim Splittorff and Jac Verbaarschot
for comments and discussions on the first version of this manuscript.


\end{document}